\documentclass[twocolumn]{aastex62}

\usepackage{savesym}
\savesymbol{tablenum}
\usepackage{siunitx}
\restoresymbol{SIX}{tablenum}

\graphicspath{{./}{figures/}}

\received{xxx}
\revised{xxx}
\accepted{xxx}
\submitjournal{ApJ}

\shorttitle{Dust Evolution in the Vicinity of a Jupiter-mass Planet}
\shortauthors{Dr\c{a}\.{z}kowska et al.}
\begin{document}

\title{Including Dust Coagulation in Hydrodynamic Models of Protoplanetary Disks: \\ Dust Evolution in the Vicinity of a Jupiter-mass Planet}

\correspondingauthor{Joanna Dr\c{a}\.{z}kowska}
\email{joanna.drazkowska@lmu.de}

\author[0000-0002-9128-0305]{Joanna Dr\c{a}\.{z}kowska}
\affil{University Observatory, Ludwig Maximilian University of Munich, Scheinerstr. 1, 81673 Munich, Germany}

\author[0000-0002-4142-3080]{Shengtai Li}
\affiliation{Theoretical Division, Los Alamos National Laboratory, Los Alamos, NM 87545, USA}

\author[0000-0002-1899-8783]{Til Birnstiel}
\affiliation{University Observatory, Ludwig Maximilian University of Munich, Scheinerstr. 1, 81673 Munich, Germany}

\author[0000-0002-1589-1796]{Sebastian M. Stammler}
\affiliation{University Observatory, Ludwig Maximilian University of Munich, Scheinerstr. 1, 81673 Munich, Germany}

\author[0000-0003-3556-6568]{Hui Li}
\affiliation{Theoretical Division, Los Alamos National Laboratory, Los Alamos, NM 87545, USA}



\begin{abstract}
 Dust growth is often neglected when building models of protoplanetary disks due to its complexity and computational expense. However, it does play a major role in shaping the evolution of protoplanetary dust and planet formation. In this paper, we present a numerical model coupling 2-D hydrodynamic evolution of a protoplanetary disk, including a Jupiter-mass planet, and dust coagulation. This is obtained by including multiple dust fluids in a single grid-based hydrodynamic simulation and solving the Smoluchowski equation for dust coagulation on top of solving for the hydrodynamic evolution. We find that fragmentation of dust aggregates trapped in a pressure bump outside of the planetary gap leads to an enhancement in density of small grains. We compare the results obtained from the full coagulation treatment to the commonly used, fixed dust size approach and to previously applied, less computationally intensive methods for including dust coagulation. We find that the full coagulation results cannot be reproduced using the fixed-size treatment, but some can be mimicked using a relatively simple method for estimating the characteristic dust size in every grid cell.
\end{abstract}



\section{Introduction} \label{sec:intro}

Due to the expense of including dust coagulation in already expensive hydrodynamic models of protoplanetary disks,  (magneto-)hydrodynamic codes usually adopt either fixed size or fixed Stokes number approach, and the size distribution is taken into account either by stacking results of series of single-sized models or including multiple dust fluids representing different dust sizes in one simulation (but without the possibility to exchange mass between the different size fluids as would happen during coagulation, see, e.g. \citealt{2010ApJ...722.1437B, 2014ApJ...785..122Z, 2015A&A...584A.110P, 2015MNRAS.453L..73D, 2015ApJ...804...35R, 2016ApJ...818...76J, 2016A&A...590A..17R, 2017ApJ...847...52X, 2017ApJ...835..118M, 2018ApJ...867....3H, 2018A&A...618A..75S}).

Dust coagulation is usually studied in azimuthally and vertically averaged setups \citep{2008A&A...480..859B, 2010A&A...513A..79B, 2012ApJ...752..106O, 2016ApJ...818..200E, 2018ApJ...868..118H, 2019ApJ...878...39L}. The dust component is typically treated as a fluid and the Smoluchowski equation is used to solve for dust coagulation. The alternative is to treat dust as (super-)particles and use the Monte Carlo approach to model collisions \citep{2007A&A...461..215O, 2008A&A...489..931Z, 2013A&A...556A..37D, 2018A&A...611A..18L, 2019ApJ...874...26S}.
A limited number of hybrid algorithms, which connect the two approaches have been developed as well \citep{2012ApJ...753..119C, 2018ApJ...864...78K}.

The connection between hydrodynamic simulations was previously done by taking azimuthally averaged profile of gas obtained in hydrodynamic simulations and performing the dust coagulation calculation in a post-processing step \citep{2012A&A...545A..81P, 2016ApJ...823...80C}.

\citet{2018ApJ...863...97T} included a simplified prescription for dust growth in hydrodynamic code RoSSBi \citep{2015A&A...579A.100S, 2016ApJ...831...82S}, where dust is represented by a single fluid but dust size may be different in every cell and is set in a sub-grid algorithm based on the work of \citet{2012A&A...539A.148B}, and demonstrated that this approach yields significantly different outcome than fixed-size treatment. \citet{2018A&A...614A..98V} implemented a similar method, with two dust populations, where dust growth is limited by barriers as proposed by \citet{2012A&A...539A.148B}. However, the method proposed by \citet{2012A&A...539A.148B} was developed for azimuthally averaged, smooth disk and was previously not tested in a full disk setup. \citet{2017MNRAS.467.1984G} included dust coagulation in a 3-D smoothed-particle hydrodynamics code (SPH), however the particle size distribution is not taken into account in the coagulation calculation (i.e. single-sized growth is modeled inside of every SPH dust super-particle).

In this paper, for the first time, we test different approaches to include dust coagulation in 2-D hydrodynamic simulations of protoplanetary disks against the fully self-consistent dust coagulation prescription included in the hydrodynamic grid code \texttt{LA-COMPASS} \citep{2005ApJ...624.1003L, 2009ApJ...690L..52L, 2014ApJ...788L..41F, 2014ApJ...795L..39F, 2019ApJ...878...39L}.

We focus on the problem of dust growth in the vicinity of a Jupiter-mass planet, which interacts with the protoplanetary disk and modifies the evolution of gas and dust. The problem of dust evolution in the neighbourhood of a gap-opening planet is particularly important for further growth of the planet by pebble accretion \citep{2018A&A...615A.110A, 2018A&A...612A..30B}. Early accretion of Jupiter in the Solar System is thought to impact the subsequent accretion of the other planets \citep{2012ApJ...756...70K, 2015A&A...582A..99I}. The growing Jupiter is also considered as a barrier for mixing of different reservoirs in the early solar nebula \citep{2017PNAS..114.6712K, 2018NatAs...2..873A}.

This paper is organized as follows. We describe our setup and numerical methods in section~\ref{sect:methods}. We describe the results in section~\ref{sect:results}. In section~\ref{sect:discussion}, we discuss the limitations and implications of the results. Finally, we summarize our work in section~\ref{sect:summary}.

\section{Methods}\label{sect:methods}

\begin{deluxetable}{ll}[tbp!]
 \tablecaption{{Input parameters of our model.} \label{tab:input}}
 \tablecolumns{2}
 \tablehead{
  \colhead{Description} &
  \colhead{Value}
 }
 \startdata
 Gas surface density at 1~AU & 1700~g~cm$^{-2}$ \\
 Gas surface density exponent & -1.5 \\
 Dust-to-gas ratio & 0.01 \\
 Temperature at 1 AU & 195~K \\
 Temperature exponent & -0.5 \\
 Turbulence strength parameter $\alpha$ & 10$^{-3}$ \\
 Planet mass & 1~M$_{\rm J}$ \\
 Planet semi-major axis & 10~AU \\
 Initial / minimum dust size & $10^{-4}$~cm \\
 Internal density of dust grains & 1.2~g~cm$^{-3}$ \\
 Dust fragmentation threshold & 10~m~s$^{-1}$ \\
 \enddata
\end{deluxetable}

We study the evolution of a protoplanetary disk around a solar mass star and follow the Minimum Mass Solar Nebula model (MMSN, \citealt{1977Ap&SS..51..153W}) characterized by the gas surface density profile
\begin{equation}
 \Sigma_{\rm g,t=0} = 1700 \cdot \left(\frac{r}{1~\textrm{AU}}\right)^{-3/2} \mathrm{g~cm}^{-2},
\end{equation}
where $r$ is the radial distance to the central star. The initial distribution of dust follows the gas profile with radially constant dust-to-gas ratio of 1\%. In models including dust coagulation, the initial size of grains is set to $a_0=1~\mu$m.

The isothermal temperature structure of the disk is set by the sound speed in the gas $c_\mathrm{s}$ and
\begin{equation}\label{eq:cs}
 c_{\mathrm s} = 83745.82 \cdot \left(\frac{r}{1~\textrm{AU}}\right)^{-1/4} \mathrm{cm~s}^{-1},
\end{equation}
which translates into $T = 195~\mathrm{K} \cdot (r/{1~\mathrm{AU}})^{-1/2}$ (we adopt the mean molecular weight of $\mu=2.3$~proton mass). The temperature profile is fixed during the simulation.
We assume that the scale-height of the disk is $H_{\mathrm g}=c_{\mathrm s}/\Omega_{\mathrm K}$, where $\Omega_{\mathrm K}$ is Keplerian frequency. With these assumptions, $c_{\mathrm s}/v_{\mathrm{K}} = H_{\mathrm g}/r \propto r^{1/4}$, so the disk is slightly flaring. We assume a gas kinematic viscosity $\nu = \alpha c_{\mathrm s} H_{\mathrm g}$ \citep{1973A&A....24..337S} and we set $\alpha=10^{-3}$. We focus on the region between 4~AU and 34~AU and place a Jupiter mass planet at a fixed, circular orbit with a semi-major axis of 10~AU. With the adopted disk parameters, $H_{\mathrm g}/r=0.05$ at the planet location.  The planet gradually opens a gap in the disk, modifying the radial distribution of both gas and dust. Table~\ref{tab:input} summarizes the input values used in all the models. A very similar setup was used in \citet{2018ApJ...863...97T}, however they assumed an inviscid disk.

In this paper, we use four different methods to model the evolution of this system of gas, planet, and dust:
\begin{itemize}
 \item {\bf Fixed size}: 2-D hydrodynamic simulations with one dust fluid representing dust of a fixed size.
 \item {\bf 1-D coagulation}: 1-D dust coagulation simulation using the azimuthally averaged gas evolution obtained from the 2-D hydrodynamic models as an input to simulate multiple 1D dust fluids to resolve dust coagulation and radial transport.
 \item {\bf Simple coagulation}: 2-D hydrodynamic simulation with one dust fluid and a sub-grid method that sets the dust size in each cell according to an expected coagulation outcome (similar to the method proposed by \citealt{2018ApJ...863...97T}).
 \item {\bf Full coagulation}: 2-D hydrodynamic simulation with multiple dust fluids representing the full size distribution, and dust coagulation algorithm which redistributes mass between the fluids.
\end{itemize}
We describe each of these methods in detail below.

\subsection{2-D models}\label{sub:2Dmodels}
All the 2-D ($r+\phi$) hydrodynamic models are performed using code which is a part of \texttt{LA-COMPASS} (which stays for Los Alamos CoMPutational AStrophysics Suite) and was described by \citet{2005ApJ...624.1003L, 2009ApJ...690L..52L}. The protoplanetary disk is assumed to be geometrically thin so that the hydrodynamical equations can be reduced to two-dimensional Navier-Stokes equations by considering vertically integrated quantities. We adopt a locally isothermal equation of state
\begin{equation}
 P_{\rm g}(r) = \Sigma_{\rm g} c_{\rm s}^2,
\end{equation}
where $P_{\rm g}$ is the vertically integrated pressure, $\Sigma_{\rm g}$ is the gas surface density and $c_{\rm s}$ is the sound speed, which only depends on distance $r$ here (see \autoref{eq:cs}).

Dust is treated as a pressureless fluid in a bi-fluid model that is governed by conservation laws \citep[see, e.g.,][]{2006A&A...453.1129P, 2012A&A...545A.134M}.  The gas and dust equations are coupled together through source terms that model the drag between the two fluids, i.e.~including the backreaction from dust to gas. We include both Epstein and Stokes drag regimes. We have implemented a Godunov Riemann solver for dust equations. We also implement dust diffusion due to the turbulence and consistently combine it with the dynamic model of bi-fluid \citep[see][]{2014ASPC..488...96L, 2014ApJ...795L..39F}. To deal with multiple timescales of coupling different dust species with gas dynamics, we develop an efficient and robust $L$-stable method to solve the coupled gas and dust equations \citep{2017AIPC.1863X0004L}.

The planet motion is governed by Newton's laws, whose equations are solved with an adaptive high-order Runge-Kutta method. Planet's gravitational potential is smoothed over 0.7 disk scale-height $H_{\mathrm g}$ at the planet location. The disk self-gravity is not included in the models presented in this paper.

We used linear polar grid with uniform spacing between the cells. The grid resolution is $N_{\mathrm r}\times N_\phi$ = $1024\times1024$. With this resolution, the disk scale-height $H_{\mathrm g}$ is resolved with 17 cells and the Hill radius of the planet is resolved with 23 cells at the planet location.

The computational domain is set from 4~AU to 34~AU. We keep the gas density constant at the inner and outer boundaries. This is justified because no significant viscous evolution is expected for the duration of the simulations ($\sim10^5$~years). For dust, we have open inner boundary to allow outflow. We keep the dust density at the outer boundary constant for the first 1000 planet orbits, which is equivalent to a steady inflow, and then we close the boundary, mimicking a decreasing flux of dust from the outer disk to the planet region. In the initial condition, velocities of gas and dust are set according to their equilibrium values derived by \citet{1986Icar...67..375N}. At the inner and outer boundaries, the velocities (both radial and azimuthal) are kept constant during the simulation.

\subsubsection{Fixed size dust}

In the default version of the code, dust is treated as a single fluid with a fixed particle size. We run a series of models covering dust sizes between 1 micron ($10^{-4}$~cm) and 10 centimeter.

\subsubsection{Full dust coagulation treatment}

The default code has been modified to include multiple dust fluids representing different dust sizes. Collisional evolution of dust is solved using explicit integration of the Smoluchowski equation. We include the Brownian motion, turbulence, differential radial and azimuthal drift, and vertical settling as sources of the collision velocities. The values of radial and azimuthal velocities for each dust species are obtained directly from the hydrodynamic solver and the other three sources are calculated in the same way as in \citet{2010A&A...513A..79B}.

When calculating the collision probabilities, we take into account midplane density of dust, which we calculate for each dust species $i$ from the surface density $\Sigma_{\rm d}^i$ used in the 2-D version of \texttt{LA-COMPASS}:
\begin{equation}\label{eq:rhod}
 \rho_{\rm d}^i = \frac{\Sigma_{\rm d}^i}{\sqrt{2\pi}H_{\rm d}^i},
\end{equation}
where we assume Gaussian distribution of the grains around the midplane with scale-height following \citet{1995Icar..114..237D}
\begin{equation}\label{eq:hd}
 H_{\rm d}^i = H_{\rm g} \cdot\sqrt{\frac{{\alpha}}{\alpha+{St^i}}},
\end{equation}
where $H_{\rm g}$ is the gas scale-height. For small grains, this equation is consistent with the work of \citet{2007Icar..192..588Y}. In this approach, we assume that turbulent mixing is fast enough to always keep the vertical structure in the settling-mixing equilibrium. This assumption might break in a low turbulence case, when the interplay between settling and dust growth leads to the so-called sedimentation-driven coagulation \citep[see, e.g.,][]{2011A&A...534A..73Z}.

We assume that grains are compact spheres with internal density of $\rho_{\rm s}=1.2$~g~cm$^{-3}$. Collisional outcomes include sticking for collisions with the impact speed below $v_{\rm f}=10$~m~s$^{-1}$, fragmentation for collisions speeds above $v_{\rm f}$, and erosion for collisions speeds above $v_{\rm f}$ when the mass ratio of colliding particles is greater than 10. Numerical implementation of the collisional evolution is the same as described in \citet[][their section 2.6]{2010A&A...513A..79B}.

The dust size distribution is resolved with 151 dust fluids covering sizes between 1~$\mu$m and 100~cm, which corresponds to 8.4 grid points per mass decade, a typical resolution used in dust coagulation models. In the initial condition, all the dust has a radius of $\SI{1}{\mu m}$.

Due to the computational expense of solving dust coagulation, we call the coagulation solver every 50 time-steps of the hydrodynamic solver. We tested that this sub-stepping routine does not impact the results significantly by running an analogical simulation where coagulation was solved at every time step (but with a shorter duration).

A more detailed description of the code will be given in a corresponding paper by Li et al. (in prep).

\subsubsection{Simple dust coagulation approach}\label{sect:simple}

We implemented a simple, sub-grid method for dust growth in the \texttt{LA-COMPASS} code, which is an updated version of the method adopted by \citet{2018ApJ...863...97T}. In this method, dust is treated as a single fluid but its size is not fixed. The dust size is calculated at every time-step and in every cell based on local conditions. In the initial condition, the size in all cells is set to $a_0 = 1~\mu$m. The initial Stokes number of grains is calculated as
\begin{equation}\label{eq:st0}
 St_0 = \frac{\pi}{2}\frac{a_0 \rho_{\rm s}}{\Sigma_{\rm g}},
\end{equation}
as all the micron-sized grains in our computational domain are in the Epstein drag regime.
Dust growth is modeled as 
\begin{equation}\label{eq:stini}
 a_i = a_{i-1} + \dot{a} \cdot \Delta t
\end{equation}
where $a_{i-1}$ is dust size obtained in the given cell in the previous time-step, $\Delta t$ is the length of the time-step, and the growth speed $\dot{a}$ is calculated as \citep[see, e.g.][]{2001A&A...378..180K}
\begin{equation}\label{eq:dotSt}
 \dot{a} = \frac{\rho_{\rm d} \Delta v}{\rho_{\rm s}},
\end{equation}
where $\rho_{\rm d}$ is the midplane density of dust (see equation \ref{eq:rhod}), $\Delta v$ is impact velocity between grains of Stokes number equivalent to $St_{i-1}$ and $0.5\cdot St_{i-1}$ (where $St_{i-1}$ corresponds to the Stokes number of grains with size $a_{i-1}$), and $\rho_{\rm s}$ is the internal density of grains. When calculating the impact speed $\Delta v$, we take into account turbulence, radial drift and azimuthal drift. The impact speeds are calculated from the radial and azimuthal velocities returned by the hydrodynamic solver assuming that the radial speed depends on the Stokes number as $v_{\rm r}\propto{St}$ (correct for ${St}<1$) and the azimuthal speed as $v_{\phi}\propto1/(1+{St}^2)$.

Dust growth can be halted either by fragmentation or radial drift. To take this into account, we calculate the maximum dust size that could grow using the semi-analytic expressions derived by \citet{2012A&A...539A.148B}.
The maximum Stokes number with turbulence-driven fragmentation is
\begin{equation}\label{eq:stfrag}
 {St}_{\rm frag} = {\rm f_f} \cdot \frac{v_{\rm f}^2}{3\alpha c_{\rm s}^2},
\end{equation}
where the fudge factor ${\rm f_f}=0.37$, and $v_{\rm f}$ is the fragmentation threshold velocity which we set to 10 m~s$^{-1}$ in this paper. This equation was derived assuming that the turbulence driven impact velocities scale as $\Delta v_{\rm t}\propto\sqrt{St}$, which applies for grains in so-called fully intermediate regime of \citet{2007A&A...461..215O}. In fact, grains which hit the fragmentation barrier are typically in this regime as shown by \citet{2011A&A...525A..11B}. 
Fragmentation can also be caused by the differential drift and the maximum Stokes number for the drift-induced fragmentation is
\begin{equation}\label{eq:stdf}
 {St}_{\rm df} = {\rm f_f} \cdot \frac{v_{\rm f}}{|\eta| v_{\rm K}},
\end{equation}
where $\eta v_{\rm K}$ is the maximum drift speed calculated using the midplane radial pressure gradient
\begin{equation}\label{eq:eta}
 \eta v_{\rm K} = \frac{1}{2 \rho_{\rm g} \Omega_{\rm K}}\cdot\frac{dP_{\rm g,z=0}}{dr},
\end{equation}
where $\rho_{\rm g}$ is the midplane gas density and $P_{\rm g, z=0}$ is midplane gas pressure.

Fragmentation is the dominant factor in setting dust size when the coagulation timescale is shorter than the drift timescale. However, in a realistic disk, this is not always true. Particularly, in the outer part of the disk, radial drift may be faster than coagulation. This sets a limit on how far the growth can proceed before the grains are removed faster that they can grow. This effect is naturally recovered in the full coagulation models, in which each dust fluid can be advected at its own speed. Although the radial drift is still accurately modeled by the hydrodynamic solver, in the simple coagulation approach the advection of dust does not have a direct effect on the representative size. Therefore we must include the drift limit explicitly in the size calculation.
The maximum Stokes number which can remain at given location taking into account radial drift is
\begin{equation}\label{eq:stdrift}
 {St}_{\rm drift} = {\rm f_d} \cdot \frac{1}{2|\eta|}\frac{\Sigma_{\rm d}}{\Sigma_{g}},
\end{equation}
where the fudge factor ${\rm f_d}=0.55$. 

The values of the fudge factors ${\rm f_f}$ and ${\rm f_d}$ that we adopted were derived by \citet{2012A&A...539A.148B} by comparing simple coagulation results to 1-D coagulation in a framework of a global, smooth disk.

The new Stokes number is decided by choosing the minimum of the values calculated when taking into account growth (${St}_i$, corresponding to the size $a_i$ obtained in equation~\ref{eq:stini}), and the possible barriers (equations \ref{eq:stfrag}, \ref{eq:stdf}, and \ref{eq:stdrift}):
\begin{equation}\label{eq:st}
 {St} = \min{\left({{St}_i},{St}_{\rm frag},{St}_{\rm df},{St}_{\rm drift} \right)}.
\end{equation}
We found that, particularly in case when pressure gradient is briefly enhanced by the spiral wakes (see figure~\ref{fig:eta}), the Stokes number recovered from this treatment can be much lower than the one given by full coagulation results. This is because the radial advection of size and the timescale needed to fragment all particles in a given cell are not taken into account. To minimize this effect, we limit the impact of fragmentation by comparing the Stokes number obtained in equation \ref{eq:st} to the Stokes number from the previous time-step ${St}_{i-1}$, and if ${St}<{St}_{i-1}$ we set
\begin{equation}\label{eq:fraglim}
 {St} = \min\left(1,{\rm f_n}\right)\cdot {St} + \max\left(0,(1-{\rm f_n})\right)\cdot{St}_{i-1},
\end{equation}
where the fudge-factor ${\rm f_n}={\Delta t}/{t_{\rm{coag}}}$ is a ratio of the simulation time-step $\Delta t$ and the coagulation timescale calculated as $t_{\rm{coag}}={a}/\dot{a}$ (see equation~\ref{eq:dotSt}). This way we avoid any sudden, local drops of the Stokes number but let it decrease gradually.
Finally we limit the minimum value of the Stokes number to the Stokes number of the smallest, micron-sized grains:
\begin{equation}
 {St} = \max{\left({St},{St_0} \right)}.
\end{equation}

The main difference between our implementation of the simple coagulation and the algorithm presented by \citet{2018ApJ...863...97T} is in the treatment of the initial dust growth phase. We changed the dust growth prescription from an exponential function implemented by \citet{2018ApJ...863...97T} to calculating the growth rate based on the local conditions (see equation \ref{eq:dotSt}). We have also introduced the limit on how much can the size decrease between two consecutive time steps (see equation~\ref{eq:fraglim}).

\subsection{1-D coagulation}

To run the azimuthally averaged models, we used the \texttt{DustPy} code. The code, developed by S.~M.~Stammler and T.~Birnstiel, is \texttt{Python}-based version of the commonly used dust coagulation code described by \citet{2010A&A...513A..79B}. It solves dust coagulation and radial surface density evolution in azimuthally and vertically averaged framework, performing implicit integration of Smoluchowski equation and advection-diffusion equation.

To test the impact of solving dust coagulation in azimuthally averaged framework, and reproduce the approach previously used to study dust coagulation in the presence of gap opening planet, we set up a model where the \texttt{DustPy} code uses azimuthally averaged gas evolution obtained in the \texttt{LA-COMPASS} simulation as an input. This is done in the following way: the azimuthally averaged output of the 2-D model is stored at every 10 planet orbits. An interpolation routine is used to generate input at time instances needed by the 1-D model. Otherwise, we use the coagulation setup with parameters given in Table~\ref{tab:input}.

\section{Results}\label{sect:results}

\subsection{Full coagulation}

\begin{figure}[tbp!]
 \centering
 \includegraphics[width=\linewidth]{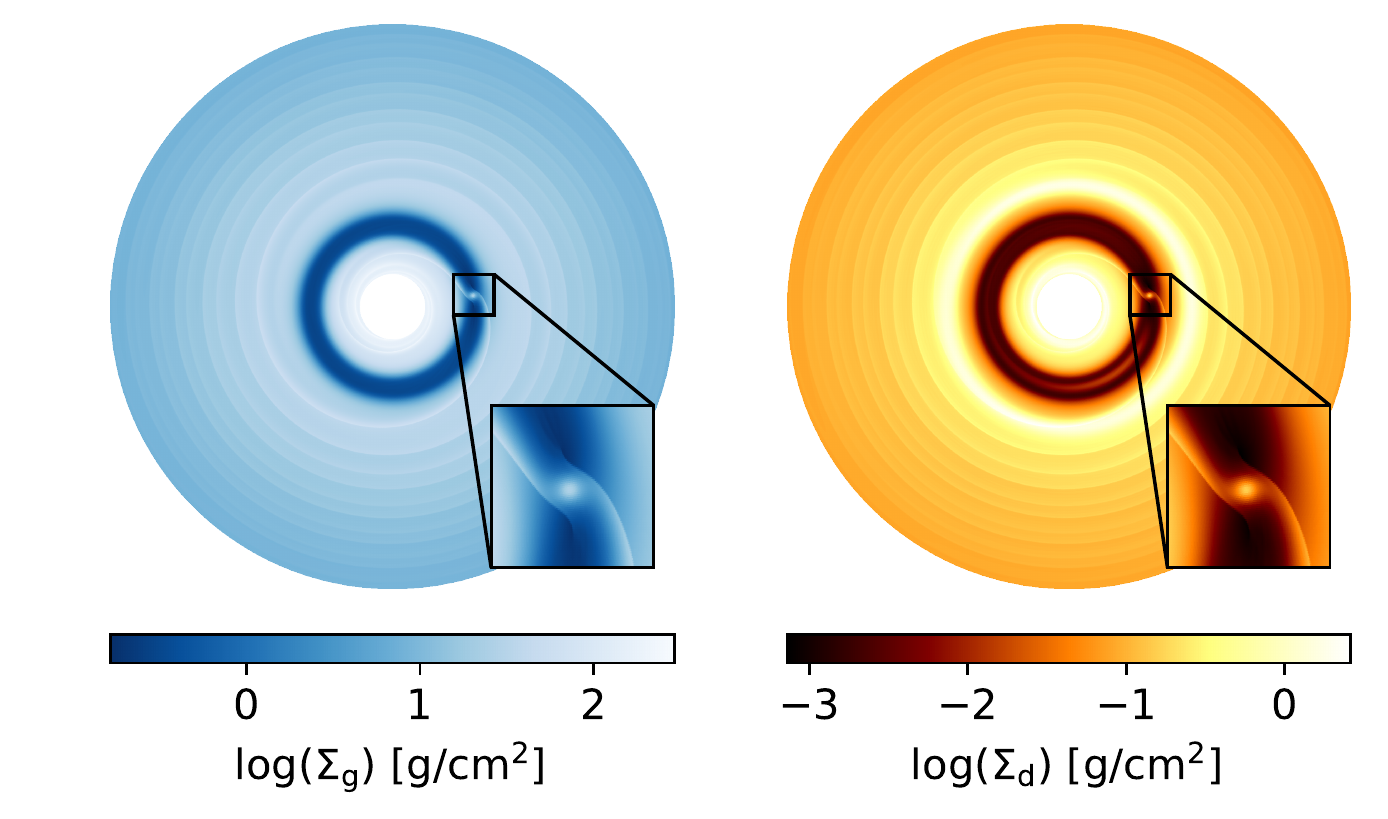}
 \caption{Surface density of gas and dust (left and right panel, respectively) obtained in the full coagulation run after 1000 planet orbits (corresponding to 31622.8 years). The inserts zoom in on the planet region.}
 \label{fig:full_coag_maps}
\end{figure}

\begin{figure}[tbp!]
 \centering
 \includegraphics[width=0.9\linewidth]{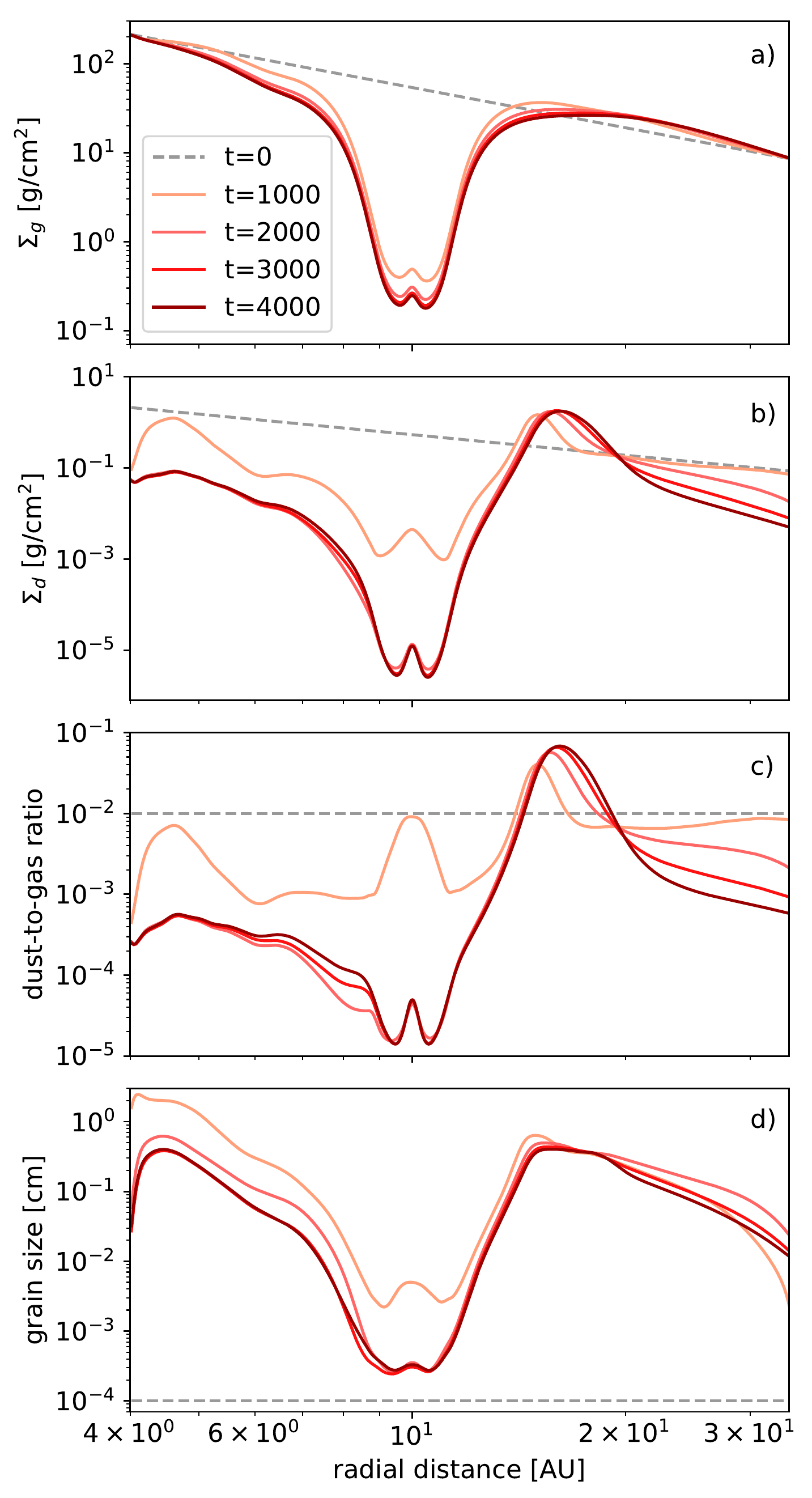}
 \caption{Azimuthally averaged time evolution of the full coagulation run. The snapshots were taken every 1000 planets orbits. {\it a)} surface density of gas, {\it b)} surface density of dust, {\it c)} vertically integrated dust-to-gas ratio, {\it d)} density averaged grain size.}
 \label{fig:full_coag_evo}
\end{figure}

As expected, the massive planet placed in a viscous disk quickly clears a gap, however some accretion flow through the gap is retained \citep{2006ApJ...641..526L}, which is visible in the Figure~\ref{fig:full_coag_maps}. The initial power-law density profiles are modified not only by planetary gap opening, but also by the planet induced spiral density waves and by dust drift. Dust evolution depends strongly on grain sizes, which influence their aerodynamic interaction with gas.

We run the simulation for 4000 planet orbits (corresponding to $t=1.26\cdot10^5$~years). Figure~\ref{fig:full_coag_evo} presents time evolution of the gas and dust surface density as well as characteristic dust size. As can be seen, the gap profile is practically saturated at the end of the simulation. As dust growth timescale is on the order of 100 local orbits (or, to be more precise, the growth timescale is orbital timescale divided by the dust-to-gas ratio, see, e.g., \citealt{2012A&A...539A.148B}), the dust sizes quickly reach a steady-state and therefore they do not change significantly after the gap is fully open.

\begin{figure}[tbp!]
 \centering
 \includegraphics[width=\linewidth]{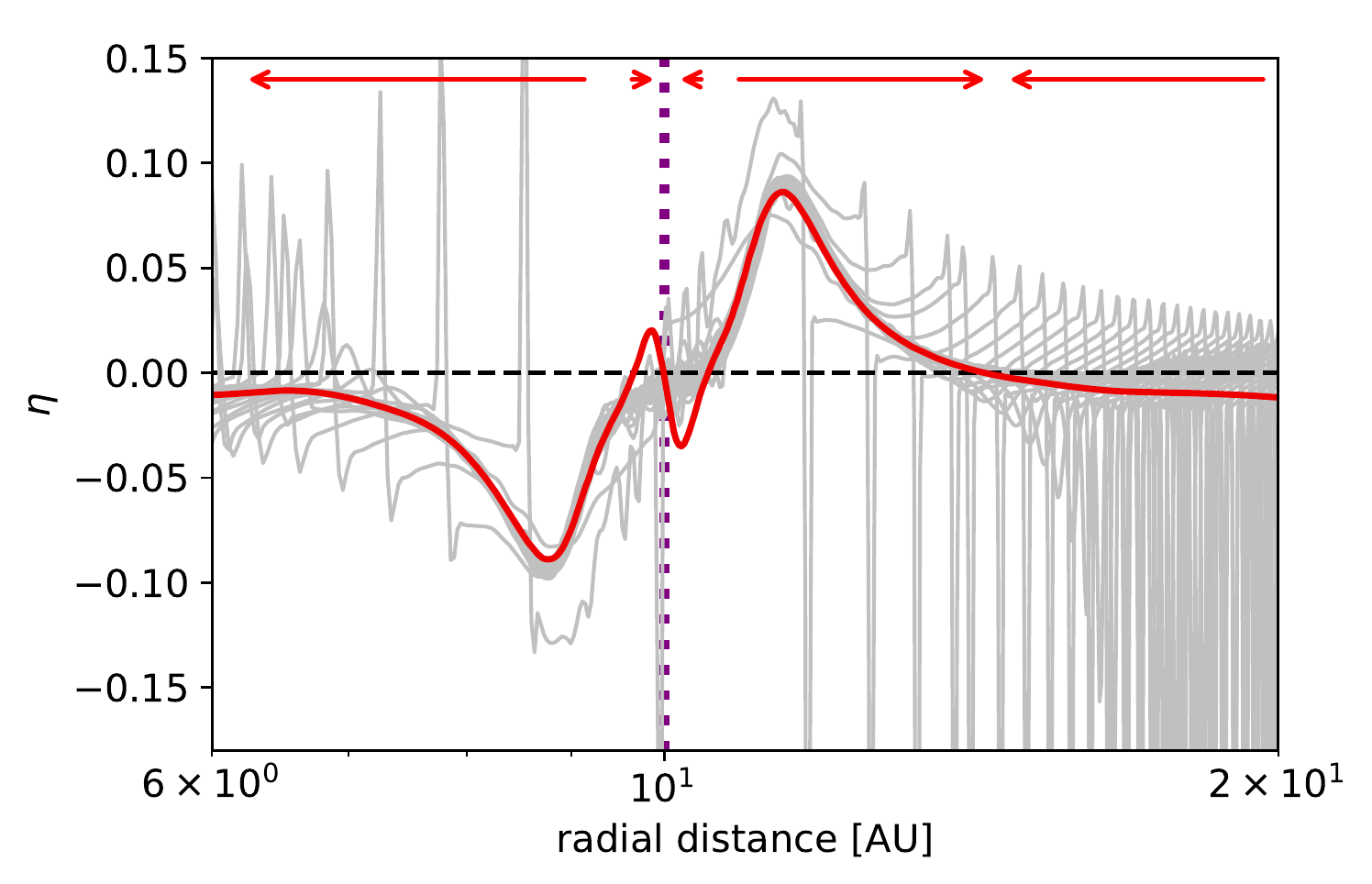}
 \caption{Radial pressure gradient parameter $\eta$ at 4000 planet orbits. The grey lines correspond to 20 azimuthal disk sectors. The red line is calculated based on azimuthally averaged gas density. The negative values of $\eta$ correspond to inward and positive values to outward dust drift (see the red arrows). The dotted vertical line marks position of the planet.}
 \label{fig:eta}
\end{figure}

Interaction between the planet and the disk causes formation of a pressure bump at the outer edge of the planet gap. Figure~\ref{fig:eta} shows the $\eta$ parameter (see equation~\ref{eq:eta}), which defines the maximum possible drift speed of dust and its direction. The red line corresponds to azimuthally averaged value of $\eta$, which determines the overall evolution of dust. Negative values of $\eta$ translate into inward drift and positive values mean outward drift of dust. As $\eta>0$ between 11~AU and 15~AU, the inward drift is reversed and thus we can expect that the radial dust drift is halted around 15~AU. Indeed, as visible in the middle panel of figure~\ref{fig:full_coag_evo}, dust is radially concentrated around 15 AU. However, trapping is not 100\% efficient and the inner region of the domain is not completely cleared, but some population of grains is retained throughout the simulation. This is because the effect of trapping is compromised by viscosity \citep[see, e.g.][]{2012A&A...545A..81P, 2018A&A...615A.110A}. Since the radial drift speed in the pressure bump is directed towards $\eta=0$ and it increases with size, there is a critical size for which the particle will always drift back to the pressure bump after being displaced by random turbulent movements. Thus, small grains are expected to pass through the gap, while large grains are expected to stay outside of the gap. Gas flows quickly through the gap and small grains, which are well-coupled, are carried along \citep{2006MNRAS.373.1619R, 2012ApJ...755....6Z}. This is confirmed in the lower panel of figure~\ref{fig:full_coag_evo}, where the density averaged size at each location is plotted. The typical size of grains in the gap is much smaller than outside of the gap. We will discuss this effect in more details in the subsequent section, where we focus on comparing the full coagulation run to models employing the fixed size approach.

It is worth noting that the radial profiles of the $\eta$ parameter plotted for different disk sectors (the grey lines in figure~\ref{fig:eta}) display significant variations, driven mostly by the spiral wakes. These wakes sweep the disk, causing temporary, small-scale pressure bumps. However, in the full coagulation approach, these do not seem to modify dust evolution considerably.

\subsection{Fixed dust size versus full coagulation}

\begin{figure*}
 \centering
 \includegraphics[width=0.75\linewidth]{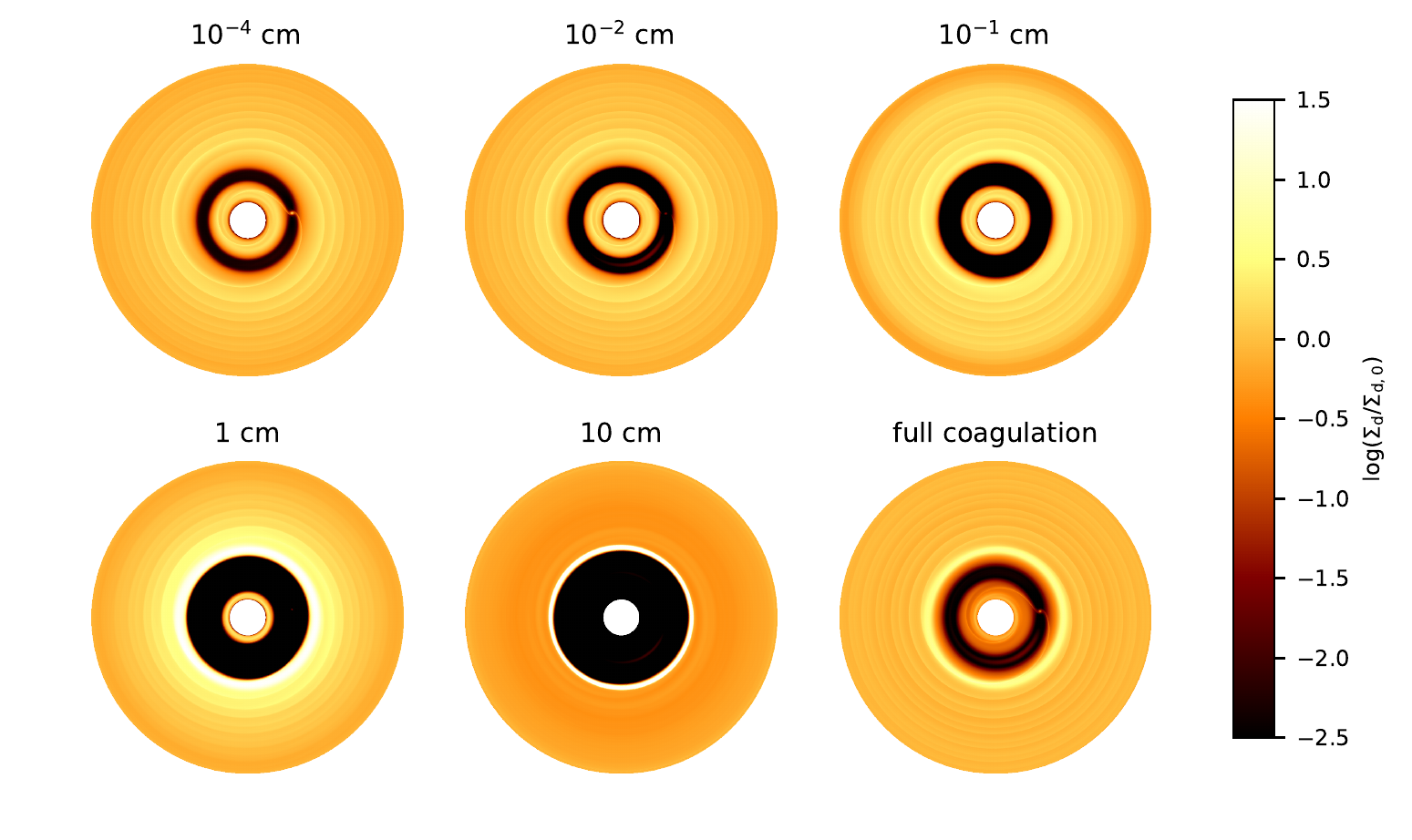}
 \caption{Snapshots of dust density obtained in the fixed size and in the full coagulation runs. Density was normalized by its initial power-law profile.  The size used in each run is indicated by panel titles. In the full coagulation run, size distribution in each cell is determined by the interplay between dust coagulation and drift.}
 \label{fig:2D_maps_coag}
\end{figure*}

\begin{figure}[tbp!]
 \centering
 \includegraphics[width=0.9\linewidth]{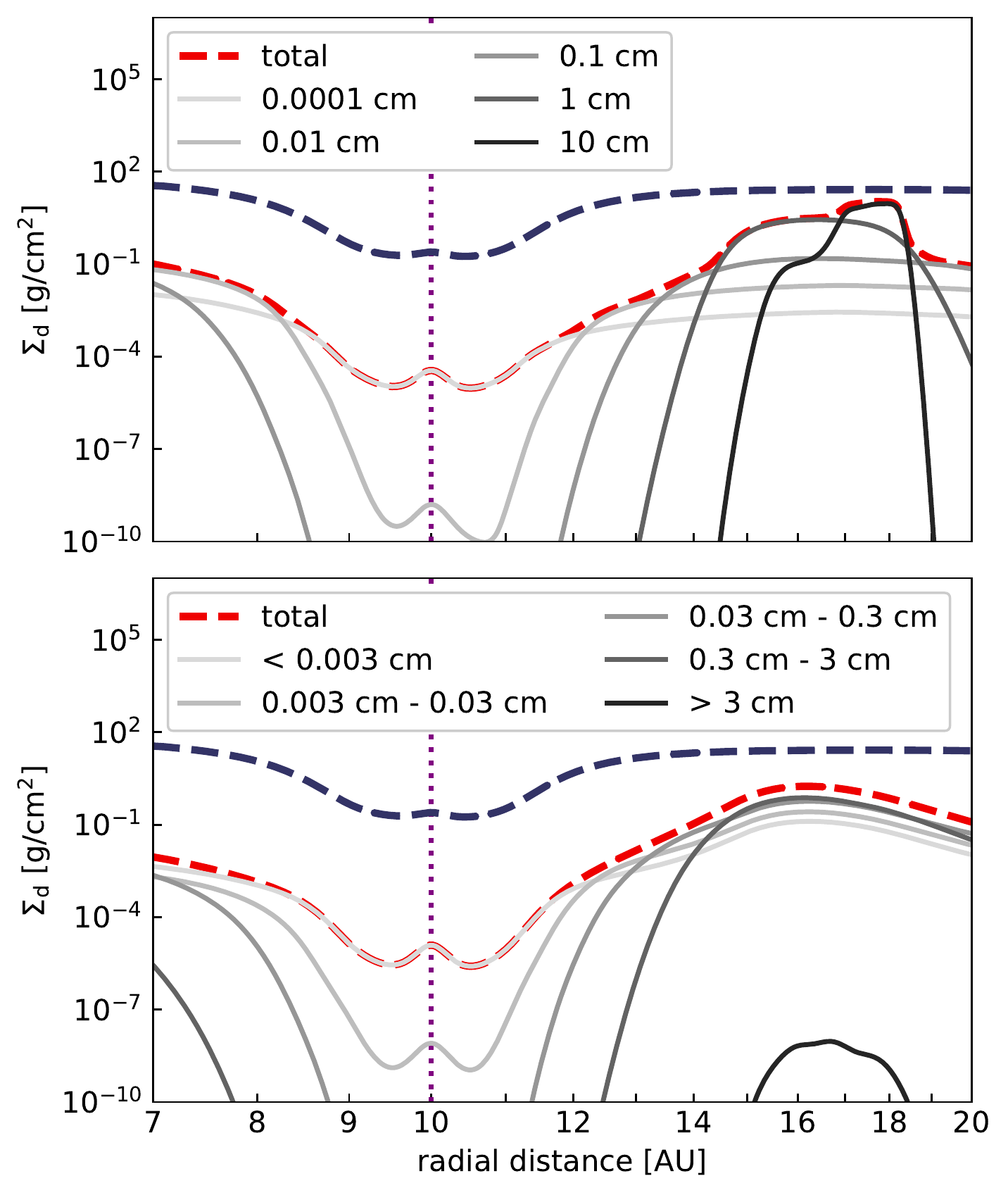}
 \caption{{\it Upper panel:} Azimuthally averaged gas (dashed blue line) and dust (solid lines) surface density profiles obtained in runs with fixed dust size after 1000 planet orbits. The red dashed line is the total dust density assuming the MRN size distribution. {\it Lower panel:} Split of the total dust density obtained in the full coagulation run (red dashed line) into contributions from different size bins (solid lines). The blue dashed line shows surface density of gas. The dotted vertical line marks the position of the planet.}
 \label{fig:filtering}
\end{figure}

Figure~\ref{fig:2D_maps_coag} compares dust distribution in the protoplanetary disk obtained in the series of models assuming fixed dust size and in the model with full coagulation.
In agreement with previously published results \citep[see, e.g.,][]{2004A&A...425L...9P, 2006MNRAS.373.1619R, 2010A&A...518A..16F, 2018ApJ...854..153W}, we find that large grains are trapped outside of the gap opened by the planet and cannot pass it. The larger the grains, the larger and deeper gap they open in dust density distribution. On the other hand, small grains that are coupled to the gas, pass through the gap. The critical size of grains that can pass trough the gap is about 1 millimeter in our setup.
The millimeter sized grains open a clear gap but at the same time do not form a distinct peak outside of the planetary gap, which is characteristic for simulations including larger dust sizes.

Figure~\ref{fig:filtering} compares the azimuthally averaged dust density profiles obtained in the series of fixed-size models (upper panel) to the results of the full coagulation model (lower panel).
We find that the results obtained when applying the full dust coagulation treatment cannot be adequately fitted using a single fixed-size model. Dust distribution resulting from the interplay between multi-size dust advection and coagulation shows both confined peak outside of the planetary gap, characteristic for centimeter-sized and larger grains, and the partially filled planetary gap, characteristic for sub-millimeter grains. Overall, the slope of dust density through the outer edge of the planet gap is much shallower in the full coagulation model than in most of the fixed-size models.

This can be understood when considering what are the contributions to the total dust density from grains of different sizes (see the lower panel of figure~\ref{fig:filtering}). While the maximum dust sizes that can be obtained in the pressure bump outside of the planet orbit are on the order of few centimeters, the gap is filled exclusively with grains smaller than 300 micron.

\begin{figure}[tbp!]
 \centering
 \includegraphics[width=0.9\linewidth]{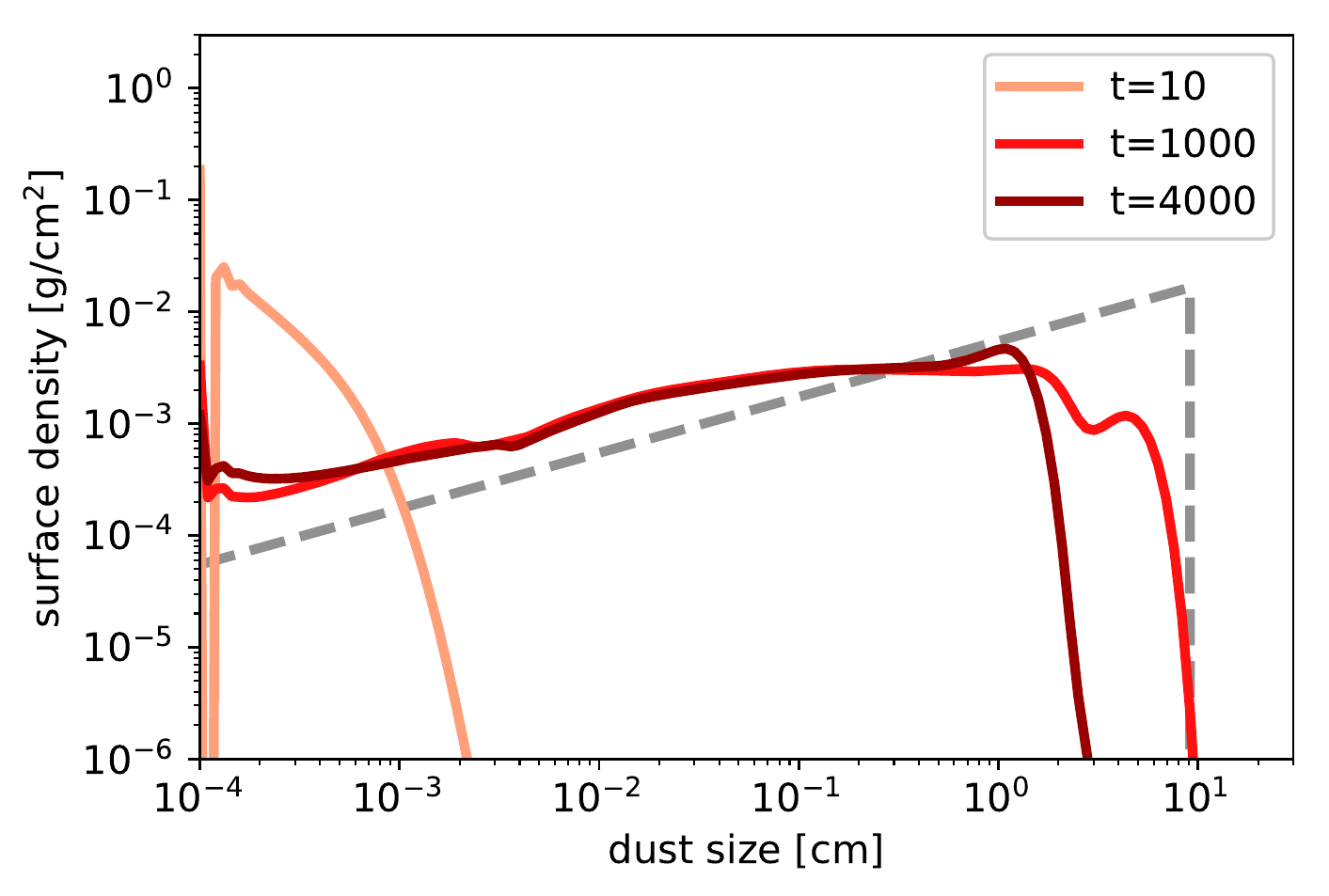}
 \caption{Comparison of the MRN size distribution with maximum size of 10~cm (dashed line) and the global dust size distribution obtained in the full coagulation model at different times (integrated over the whole disk, solid lines).}
 \label{fig:sizedistr_MRN}
\end{figure}

In models that do not solve dust coagulation but need input on size distribution (mostly for comparison to observations), it is often assumed that the dust size distribution follows the so called MRN distribution with $n(a)\propto a^{-3.5}$ \citep{1977ApJ...217..425M}. In the upper panel of figure~\ref{fig:filtering}, we summed up contributions from each single-size model assuming the MRN distribution. The resulting total density profile (red dashed line) is relatively similar to the density profile obtained in the full coagulation model presented in the bottom panel. In figure~\ref{fig:sizedistr_MRN} we compare the global size distribution obtained in the full models to the assumed MRN profile. At 1000 orbits of the planet, it does indeed match the power-law distribution reasonably well, although the slope of the distribution is generally shallower and there is significantly less of the largest grains. The MRN profile may be a reasonable assumption for the overall size distribution, although a good estimate of the maximum dust size is necessary, as most of the mass is contained in the largest grains. The size distribution is significantly different at the beginning of the simulation, when the grains have not reached their maximum sizes yet. Also, the maximum size of grains decreases toward the end of the simulation, and falls from 10~cm to about 3~cm at 4000 planet orbits. This is caused by the decreasing dust influx at the outer boundary (see Section \ref{sub:2Dmodels}). With a lower dust-to-gas ratio, the drift barrier affects smaller grains \citep{2012A&A...539A.148B}. Thus, the MRN size distribution with a fixed maximum dust size may not be very useful to model disk evolution over a long period of time.

\begin{figure}[tbp!]
 \centering
 \includegraphics[width=0.9\linewidth]{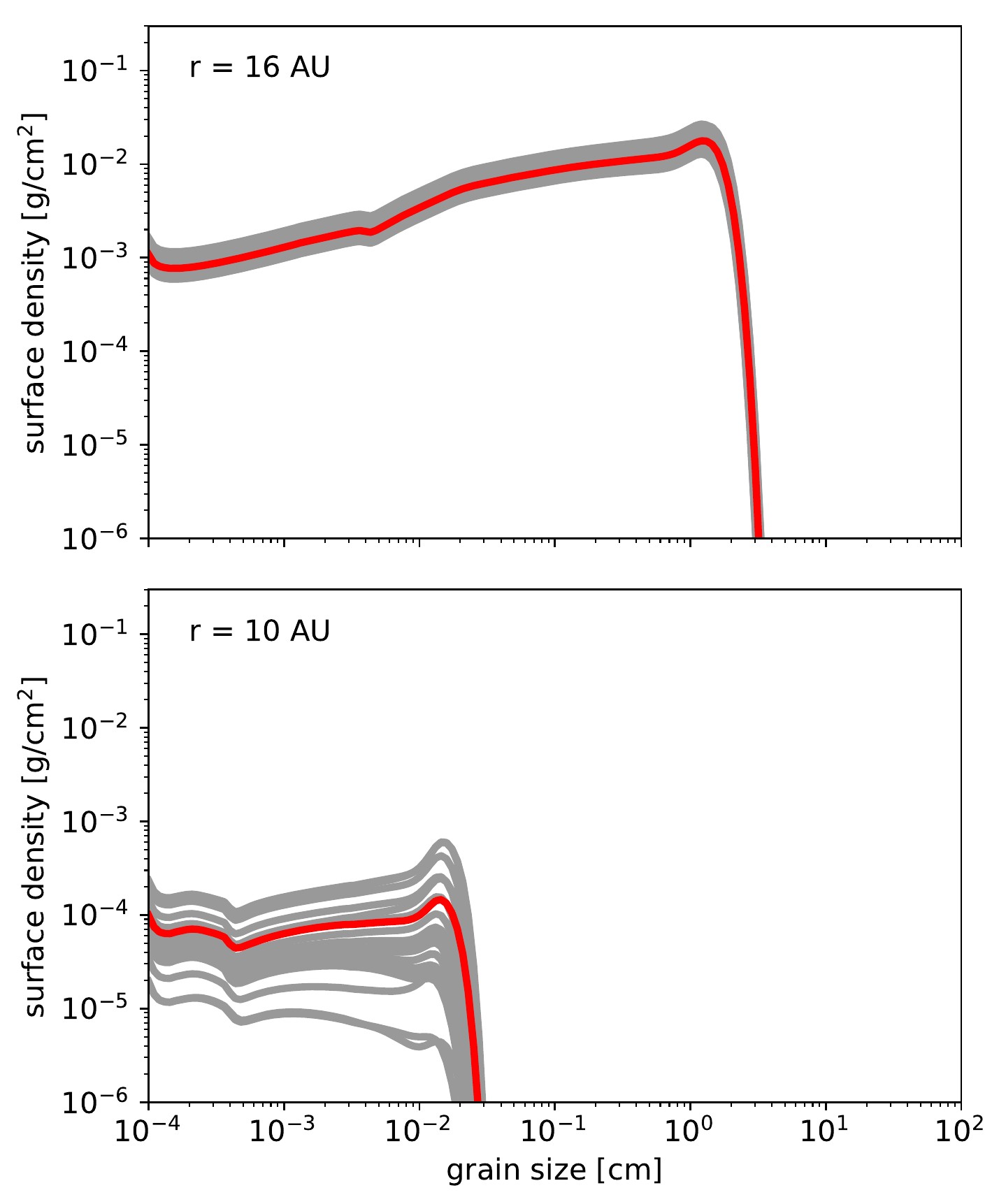}
 \caption{Dust size distributions obtained at 10 AU ({\it lower panel}) and at 16~AU ({\it upper panel}) in the full coagulation model. The red line corresponds to azimuthally averaged profile while the gray lines represent sample distributions across 20 homogeneously distributed angles.}
 \label{fig:sizedistr}
\end{figure}

A significant difference between the full coagulation and fixed-size models is that in the full models fragmentation constantly replenishes small grains that can pass through the planetary gap. The impact of fragmentation is noticeable when comparing the upper and lower panels of figure~\ref{fig:filtering}. In the full coagulation simulation (lower panel), the small grains follow the density profile of larger grains outside of the planet orbit, different than expected from the fixed-size simulations (upper panel). This is because the small grains are constantly produced in collisions between larger aggregates. Our results suggest that, if fragmentation is efficient, density of small aggregates should be enhanced in dust traps, despite they are not trapped themselves. Indeed, a look into the size distribution of grains presented in figure~\ref{fig:sizedistr} reveals that outside of the planet orbit, the size distribution is close to coagulation-fragmentation equilibrium \citep[see, e.g.][]{2011A&A...525A..11B}.

Another effect that distinguishes the full-coagulation simulation from the fixed-size models is the growth of small particles after they passed the gap. Dust density in the gap is too low to allow for efficient coagulation. But in the inner region of the simulation domain, where dust density increases, grains larger than 3 centimeters are present, which would not be expected from the fixed-size models.

It is worth noting that the dust distribution obtained outside of the planet gap (the upper panel of figure~\ref{fig:sizedistr}) is remarkably symmetric, with little deviation when considering different azimuthal angles. This is not true in the planet co-orbital region (lower panel of figure~\ref{fig:sizedistr}). Some of the small grains that pass through the gap are trapped either in direct vicinity of the planet (we do not consider accretion onto the planet) or in the Lagrange points (this is visible in figure~\ref{fig:full_coag_maps}). Due to this asymmetric nature of the co-orbital region, the size distributions sampled around the planet orbit exhibit different profiles.

Since the density and size distribution profiles are generally symmetric (outside of the planetary gap region) and the size distribution generally follows the expected coagulation-fragmentation equilibrium profile, we can expect that it would be possible to recover results of the full coagulation model using less computationally expensive methods. In the subsequent section, we compare results of three methods of including dust coagulation.

\subsection{Comparison of different treatments for dust coagulation}

We aimed to reproduce the results of the full coagulation run using two less computationally intensive methods. The first method relies on semi-analytical estimate of the coagulation outcome and is based on the work of \citet{2012A&A...539A.148B}. The second method relies on extracting the gas evolution from the full, 2-D simulation and running 1-D, azimuthally averaged dust evolution model in a post-processing step. The estimated computational expense of the three methods, per 1000 planet orbits, is as follows:
\begin{itemize}
 \item {Full coagulation:} 27 684 CPU hours (12 hours on 2304 CPUs).
 \item {Simple coagulation:} 192 CPU hours (20 minutes on 576 CPUs, very similar to an analogical fixed-size simulation).
 \item {1-D coagulation:} 78 CPU hours.
\end{itemize}

\begin{figure}[tbp!]
 \centering
 \includegraphics[width=0.9\linewidth]{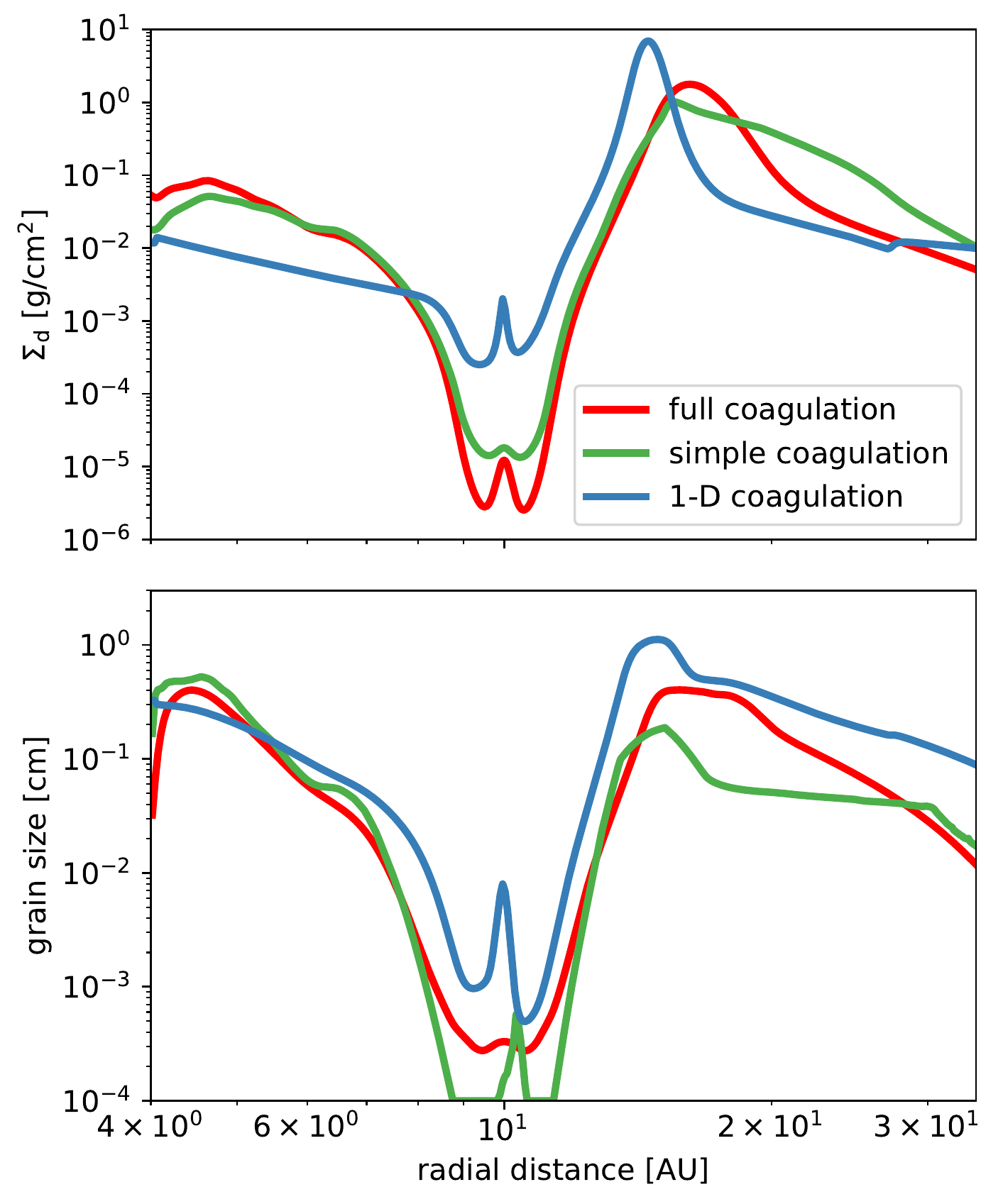}
 \caption{{\it Upper panel:} comparison of azimuthally averaged dust surface density profiles obtained in runs with full coagulation, simple coagulation, and 1-D coagulation after 4000 planet orbits. {\it Lower panel:} azimuthally averaged profiles of dust size obtained in the three simulations.}
 \label{fig:coag_lineplots}
\end{figure}

%Figure~\ref{fig:coag_2D} compares snapshots of dust densities and sizes obtained using the three methods.
Figure~\ref{fig:coag_lineplots} compares azimuthally averaged dust densities and dust sizes obtained using the three methods.
As can be inferred from this plot, the simple coagulation method is generally better in reproducing the full results than the azimuthally averaged approach.

The main problem of 1-D coagulation is that azimuthal averaging of gas density "kills" the effects of spiral wakes (see figure~\ref{fig:eta}): they induce additional impact speeds, limiting the maximum size possible to obtain, but they also induce extra mixing.
The 1-D coagulation predicts almost one order of magnitude higher peak density in the trap outside of the planet orbit. The peak is also narrower than in full coagulation results. This effect has multiple reasons: first of all, due to averaging out of the spiral wakes, the 1-D coagulation predicts larger particles, that are trapped more efficiently. On top of that, in the full coagulation results, the exact position of the trap may change with the azimuthal angle because of the asymmetric nature of planet-disk interaction. Thus the trap is radially "smeared" in the full coagulation results. Additionally, as we mentioned before, the spiral wakes induce additional mixing which is not taken into account in the 1-D model. The full coagulation model includes the effect of backreaction of dust  to gas, which additionally increase the width of dust ring (although this is not a significant contribution in our setup, see section~\ref{sect:br}). It is worth noting that similar results of widening the peak in 2-D versus 1-D models was found by \citet{2018ApJ...854..153W} for fixed size grains.

The negligence of the 2-D effects of planet-disk interaction has the most significant outcome in the planetary gap region. The 1-D coagulation predicts significantly more dust inside of the gap than the full and simple coagulation models. In 2-D simulations, dust can only flow through the gap if it enters the streamline around the plane (see figure~\ref{fig:full_coag_maps}). The 1-D model cannot take this subtlety into account. In our case, using the azimuthally averaged gas density information to calculate pressure gradient and drift speed leads to an increased dust flux through the gap. This is an opposite conclusion than presented by \citet{2018ApJ...854..153W} who also compared a 2-D and 1-D approach to dust dynamics in the vicinity of Jupiter-mass planet, using a fixed-size model. However, they did not extract the density information from 2-D simulation but run a self-consistent 1-D model with planet, which led to a different density profile in 1-D and 2-D runs.

The simple coagulation results do not reproduce the full coagulation perfectly either. The main problem of the simple model is that the size is calculated locally, without an input from neighbouring cells, thus the effect of dust mixing is not taken into account in all aspects. As shown by \citet{2012A&A...539A.148B}, the simple method reproduces results of full coagulation very well in the case a smooth, axisymmetric disk. We find that including a massive planet that induces spiral wakes, locally enhancing the pressure gradient and thus impact speeds, leads to violent fragmentation events. In the current simple coagulation approach, we only track one "representative" particle size per cell. If this size suddenly drops due to the spiral wake induced fragmentation, dust will take relatively long time to re-grow at this position. In the full coagulation method, this effect is significantly reduced as the fluids representing different sizes mix, leading to a similar size distribution in neighbouring cells. This is why we had to limit the effect of fragmentation in the simple coagulation model (see section~\ref{sect:simple}).

Despite these difficulties, the simple model reproduces the full coagulation results reasonably well. Outside of the region where the spiral wakes have the strongest effect ($\sim$10~AU to 25~AU), the dust size calculated in the sub-grid method fits the density averaged size obtained in the full coagulation model almost perfectly (see the lower panel of figure~\ref{fig:coag_lineplots}). It is worth noting that the implementation of the simple coagulation practically does not increase the computational cost of the 2-D hydrodynamic model, so this calculation is as fast as a fixed-size run.

\section{Discussion}\label{sect:discussion}

\subsection{Limitations}

We presented results of computational models utilizing state-of-the-art methods for modeling dust evolution in protoplanetary disk. However, our models are not free from limitations, which we discuss in this section.

We performed 2-D models, solving for radial and azimuthal structure of the protoplanetary disk, assuming that the vertical density distribution is Gaussian, and depends on dust size in a simple way (see equations \ref{eq:rhod} and \ref{eq:hd}). Thus we neglect potential effect that sedimentation-driven coagulation could have on dust growth \citep{2011A&A...534A..73Z}. It is known that in some cases, the 3-D effects may change the conclusions of 2-D hydrodynamic models \citep[see, e.g.,][]{2018RNAAS...2d.195L}.

We adopted a simple, isothermal protoplanetary disk model with a fixed temperature structure. Thus we do not take into account the effects of planet heating the protoplanetary disk \citep{2015MNRAS.453.1768P, 2017ApJ...842..103S}, which could potentially change the outcome of dust coagulation as the collisional speeds are dependent on the sound speed (see equation \ref{eq:stfrag}).

Our computational domain covers a patch of the protoplanetary disk ranging from 4~AU and 34~AU. While this domain allows us to cover most of the physics connected to dust drift, it is still a relatively small fraction of the global disk, which could extend to several hundreds AU. One of the problems associated with not including the outer parts of the protoplanetary disk directly is that, without a proper boundary condition, we would quickly "run out" of dust. In the models presented in this paper, we adopted an outer boundary condition which allows inflow of gas and dust, thus preventing the density at the outer edge from dropping significantly. However, particularly for a long runtime of the simulation, this condition cannot adequately account for an evolving pebble flux that is expected from global disk simulations \citep[see, e.g.,][]{2012A&A...539A.148B}.

In the full coagulation and 1-D coagulation models, we adopted a relatively simple collision model with only two possible collision outcomes: sticking and fragmentation. We have assumed a single fragmentation threshold value in the whole domain ($v_{\rm f}=10$~m~s$^{-1}$). While more complex collision models can be developed based on results of laboratory experiments \citep[see, e.g.,][]{2018SSRv..214...52B}, these are much harder to implement in the Smoluchowski equation solver \citep{2012A&A...540A..73W}. We have also neglected the evolution of porosity of dust aggregates, which can potentially lead to a different coagulation pattern \citep{2007A&A...461..215O, 2012ApJ...752..106O, 2016A&A...586A..20K}.

\subsection{Dust filtering and pebble isolation mass}

Despite these limitations, our results may have implications for the theory of pebble isolation mass. This concepts assumes that delivery of solids to a growing gap-opening planets is halted if grains are large enough to be trapped in the pressure maximum outside of the planet orbit  \citep[see, e.g.][]{2014A&A...572A..35L, 2018A&A...615A.110A, 2018A&A...612A..30B}. However, our results suggest that those large grains will fragment and constantly replenish the population of small grains, which are able to pass through the gap and potentially re-grow in the planet co-orbital region (although the resolution of our models does not allow us to resolve the potential circumplanetary disk, in which the growth would be most efficient, see \citealt{2017ApJ...846...81S, 2018ApJ...866..142D}). Dust coagulation and fragmentation could thus increase the pebble isolation mass.

Similarly, our results cast doubt on the efficiency of dust filtration by growing Jupiter which is postulated to explain some features of the Solar System. Efficient isolation of different reservoirs by a gap-opening planet, as postulated by, e.g.,~\citet{2017PNAS..114.6712K, 2018NatAs...2..873A, 2019arXiv190312274H} would only be possible if particles do not fragment during collisions (but at the same time are large enough to undergo efficient trapping). Because of the efficient fragmentation outside of the planet, small grains passing the gap around growing Jupiter could still transport water into the inner Solar System, in contrast to the idea proposed by \citet{2016Icar..267..368M}, where the proto-Jupiter blocks the delivery of water when it reaches mass of about 20 M$_{\oplus}$.

\subsection{Importantce of backreaction}\label{sect:br}

\begin{figure}[tbp!]
 \centering
 \includegraphics[width=0.9\linewidth]{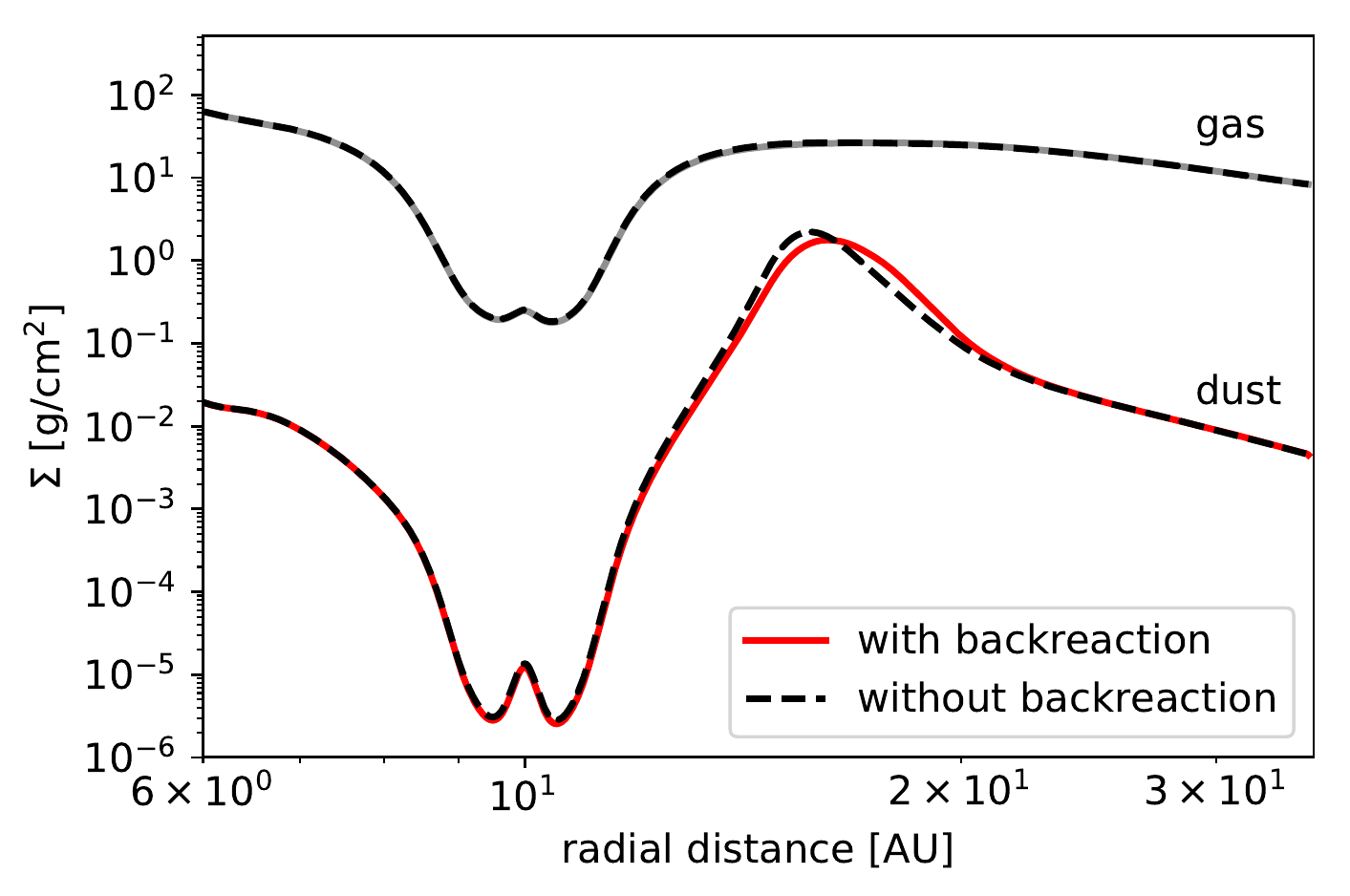}
 \caption{Azimuthally averaged surface densities of gas and dust obtained after 4000 planet orbits in the run with (solid lines) and without (dashed lines) the effect of backreaction of dust on gas included.}
 \label{fig:br}
\end{figure}

The 2-D hydrodynamic models include the effect of backreaction of dust on gas. We tested the importance of this effect by running a setup analogical to the full coagulation run but with backreaction switched off. The comparison of these two runs is presented in figure~\ref{fig:br}. The gas density is not modified significantly, but the effect of including backreaction is visible in the dust distribution. The dust ring formed outside of the planetary gap is placed a little bit further away and it is slightly wider in the run including backreaction. This is because the large grains in overdense region push on the gas, leading to a slight modification of the pressure gradient. The outward drift of dust in the pressure bump is sped up, leading to the wider ring profile. This effect was also observed by \citet{2018ApJ...868...48K}. \citet{2015MNRAS.454L..36G} suggested that backreaction may lead to formation of a second pressure maximum and, in consequence, second dust ring caused by a single planet. We do not observe such an effect in our results, but this may be due to difference in setup: our planet is significantly less massive than 5~M$_{\rm J}$ implemented by \citet{2015MNRAS.454L..36G}.

The limited effect of backreaction we observe is a consequence of assuming a viscous disk with $\alpha=10^{-3}$. Viscosity prevents the dust-to-gas ratio from becoming high: in the full coagulation model, the maximum vertically integrated dust-to-gas ratio stays below 10\% (see figure~\ref{fig:full_coag_evo}). In disks with lower viscosity, planet-disk interactions lead to development of a vortex outside of the planet orbit \citep{2009ApJ...690L..52L, 2014ApJ...788L..41F, 2017MNRAS.466.3533H}. The vortices are able to significantly concentrate dust and the effects of backreaction are more pronounced, including destruction of the vortex \citep{2014ApJ...795L..39F, 2015ApJ...804...35R, 2015MNRAS.450.4285C, 2016ApJ...831...82S}, although this effect is mitigated in 3-D models \citep{2018RNAAS...2d.195L}. We plan to study effects of dust coagulation inside of a vortex in a next paper.

\section{Summary}\label{sect:summary}

Dust coagulation is the first step toward forming planetesimals and planets. In this paper, we presented results of coupling dust coagulation to hydrodynamics in simulation of a protoplanetary disk including a massive planet. We compared our model to the usual, fixed-size approach and showed that the results differ considerably. We have also compared the full coagulation results to previously used, azimuthally averaged approach and to a simple, sub-grid growth prescription. The main findings of this work may be summarized in the following points:
\begin{itemize}
 \item Stacking fixed size simulations cannot reproduce the full coagulation results as it does not take into account the exchange of mass between dust populations of different sizes. Particularly, the fragmentation of large grains leads to enhanced density of small grains in the trap region, while the fixed size simulation does not predict trapping of small grains.
 \item Fragmentation of large grains limits the effect of trapping and increases the permeability of planet-induced gap.
 \item None of the cheaper methods of solving dust coagulation that we tested is able to recover the full coagulation result perfectly. However, both methods give a reasonable estimate of dust sizes. Any of the two methods is better in reproducing the dust density evolution than a single fixed-size approach.
\end{itemize}




\acknowledgments
We thank the referee for their valuable comments. J.D., T.B., and S.M.S. acknowledge funding from the European Research Council (ERC) under the European Unions Horizon 2020 research and innovation programme under grant agreement No. 714769 and the support from the DFG Research Unit ``Transition Disks'' (FOR 2634/1, ER 685/8-1). S.L. and H.L. gratefully acknowledge the support by LANL/CSES and NASA/ATP. Part of this work was performed at the Aspen Center for Physics, which is supported by National Science Foundation grant PHY-1607611. This research used resources provided by the Los Alamos National Laboratory Institutional Computing Program, which is supported by the U.S. Department of Energy National Nuclear Security Administration under Contract No. 89233218CNA000001.

\bibliographystyle{aasjournal}
\bibliography{paper}

\begin{thebibliography}{}
\expandafter\ifx\csname natexlab\endcsname\relax\def\natexlab#1{#1}\fi
\providecommand{\url}[1]{\href{#1}{#1}}

\bibitem[{{Alibert} {et~al.}(2018){Alibert}, {Venturini}, {Helled}, {Ataiee},
  {Burn}, {Senecal}, {Benz}, {Mayer}, {Mordasini}, {Quanz}, \&
  {Sch{\"o}nb{\"a}chler}}]{2018NatAs...2..873A}
{Alibert}, Y., {Venturini}, J., {Helled}, R., {et~al.} 2018, Nature Astronomy,
  2, 873

\bibitem[{{Ataiee} {et~al.}(2018){Ataiee}, {Baruteau}, {Alibert}, \&
  {Benz}}]{2018A&A...615A.110A}
{Ataiee}, S., {Baruteau}, C., {Alibert}, Y., \& {Benz}, W. 2018, \aap, 615,
  A110

\bibitem[{{Bai} \& {Stone}(2010)}]{2010ApJ...722.1437B}
{Bai}, X.-N., \& {Stone}, J.~M. 2010, \apj, 722, 1437

\bibitem[{{Birnstiel} {et~al.}(2010){Birnstiel}, {Dullemond}, \&
  {Brauer}}]{2010A&A...513A..79B}
{Birnstiel}, T., {Dullemond}, C.~P., \& {Brauer}, F. 2010, \aap, 513, A79

\bibitem[{{Birnstiel} {et~al.}(2012){Birnstiel}, {Klahr}, \&
  {Ercolano}}]{2012A&A...539A.148B}
{Birnstiel}, T., {Klahr}, H., \& {Ercolano}, B. 2012, \aap, 539, A148

\bibitem[{{Birnstiel} {et~al.}(2011){Birnstiel}, {Ormel}, \&
  {Dullemond}}]{2011A&A...525A..11B}
{Birnstiel}, T., {Ormel}, C.~W., \& {Dullemond}, C.~P. 2011, \aap, 525, A11

\bibitem[{{Bitsch} {et~al.}(2018){Bitsch}, {Morbidelli}, {Johansen}, {Lega},
  {Lambrechts}, \& {Crida}}]{2018A&A...612A..30B}
{Bitsch}, B., {Morbidelli}, A., {Johansen}, A., {et~al.} 2018, \aap, 612, A30

\bibitem[{{Blum}(2018)}]{2018SSRv..214...52B}
{Blum}, J. 2018, \ssr, 214, 52

\bibitem[{{Brauer} {et~al.}(2008){Brauer}, {Dullemond}, \&
  {Henning}}]{2008A&A...480..859B}
{Brauer}, F., {Dullemond}, C.~P., \& {Henning}, T. 2008, \aap, 480, 859

\bibitem[{{Carballido} {et~al.}(2016){Carballido}, {Matthews}, \&
  {Hyde}}]{2016ApJ...823...80C}
{Carballido}, A., {Matthews}, L.~S., \& {Hyde}, T.~W. 2016, \apj, 823, 80

\bibitem[{{Charnoz} \& {Taillifet}(2012)}]{2012ApJ...753..119C}
{Charnoz}, S., \& {Taillifet}, E. 2012, \apj, 753, 119

\bibitem[{{Crnkovic-Rubsamen} {et~al.}(2015){Crnkovic-Rubsamen}, {Zhu}, \&
  {Stone}}]{2015MNRAS.450.4285C}
{Crnkovic-Rubsamen}, I., {Zhu}, Z., \& {Stone}, J.~M. 2015, Monthly Notices of
  the Royal Astronomical Society, 450, 4285

\bibitem[{{Dipierro} {et~al.}(2015){Dipierro}, {Price}, {Laibe}, {Hirsh},
  {Cerioli}, \& {Lodato}}]{2015MNRAS.453L..73D}
{Dipierro}, G., {Price}, D., {Laibe}, G., {et~al.} 2015, \mnras, 453, L73

\bibitem[{{Dr{\c a}{\.z}kowska} \& {Szul{\'a}gyi}(2018)}]{2018ApJ...866..142D}
{Dr{\c a}{\.z}kowska}, J., \& {Szul{\'a}gyi}, J. 2018, \apj, 866, 142

\bibitem[{{Dr{\c a}{\.z}kowska} {et~al.}(2013){Dr{\c a}{\.z}kowska},
  {Windmark}, \& {Dullemond}}]{2013A&A...556A..37D}
{Dr{\c a}{\.z}kowska}, J., {Windmark}, F., \& {Dullemond}, C.~P. 2013, \aap,
  556, A37

\bibitem[{{Dubrulle} {et~al.}(1995){Dubrulle}, {Morfill}, \&
  {Sterzik}}]{1995Icar..114..237D}
{Dubrulle}, B., {Morfill}, G., \& {Sterzik}, M. 1995, \icarus, 114, 237

\bibitem[{{Estrada} {et~al.}(2016){Estrada}, {Cuzzi}, \&
  {Morgan}}]{2016ApJ...818..200E}
{Estrada}, P.~R., {Cuzzi}, J.~N., \& {Morgan}, D.~A. 2016, \apj, 818, 200

\bibitem[{{Fouchet} {et~al.}(2010){Fouchet}, {Gonzalez}, \&
  {Maddison}}]{2010A&A...518A..16F}
{Fouchet}, L., {Gonzalez}, J.-F., \& {Maddison}, S.~T. 2010, \aap, 518, A16

\bibitem[{{Fu} {et~al.}(2014{\natexlab{a}}){Fu}, {Li}, {Lubow}, \&
  {Li}}]{2014ApJ...788L..41F}
{Fu}, W., {Li}, H., {Lubow}, S., \& {Li}, S. 2014{\natexlab{a}}, \apjl, 788,
  L41

\bibitem[{{Fu} {et~al.}(2014{\natexlab{b}}){Fu}, {Li}, {Lubow}, {Li}, \&
  {Liang}}]{2014ApJ...795L..39F}
{Fu}, W., {Li}, H., {Lubow}, S., {Li}, S., \& {Liang}, E. 2014{\natexlab{b}},
  \apjl, 795, L39

\bibitem[{{Gonzalez} {et~al.}(2017){Gonzalez}, {Laibe}, \&
  {Maddison}}]{2017MNRAS.467.1984G}
{Gonzalez}, J.-F., {Laibe}, G., \& {Maddison}, S.~T. 2017, \mnras, 467, 1984

\bibitem[{{Gonzalez} {et~al.}(2015){Gonzalez}, {Laibe}, {Maddison}, {Pinte}, \&
  {M{\'e}nard}}]{2015MNRAS.454L..36G}
{Gonzalez}, J.-F., {Laibe}, G., {Maddison}, S.~T., {Pinte}, C., \&
  {M{\'e}nard}, F. 2015, \mnras, 454, L36

\bibitem[{{Hammer} {et~al.}(2017){Hammer}, {Kratter}, \&
  {Lin}}]{2017MNRAS.466.3533H}
{Hammer}, M., {Kratter}, K.~M., \& {Lin}, M.-K. 2017, \mnras, 466, 3533

\bibitem[{{Haugb{\o}lle} {et~al.}(2019){Haugb{\o}lle}, {Weber}, {Wielandt},
  {Ben{\'{\i}}tez-Llambay}, {Bizzarro}, {Gressel}, \&
  {Pessah}}]{2019arXiv190312274H}
{Haugb{\o}lle}, T., {Weber}, P., {Wielandt}, D.~P., {et~al.} 2019, arXiv
  e-prints, arXiv:1903.12274

\bibitem[{{Homma} \& {Nakamoto}(2018)}]{2018ApJ...868..118H}
{Homma}, K., \& {Nakamoto}, T. 2018, \apj, 868, 118

\bibitem[{{Huang} {et~al.}(2018){Huang}, {Isella}, {Li}, {Li}, \&
  {Ji}}]{2018ApJ...867....3H}
{Huang}, P., {Isella}, A., {Li}, H., {Li}, S., \& {Ji}, J. 2018, \apj, 867, 3

\bibitem[{{Izidoro} {et~al.}(2015){Izidoro}, {Morbidelli}, {Raymond},
  {Hersant}, \& {Pierens}}]{2015A&A...582A..99I}
{Izidoro}, A., {Morbidelli}, A., {Raymond}, S.~N., {Hersant}, F., \& {Pierens},
  A. 2015, \aap, 582, A99

\bibitem[{{Jin} {et~al.}(2016){Jin}, {Li}, {Isella}, {Li}, \&
  {Ji}}]{2016ApJ...818...76J}
{Jin}, S., {Li}, S., {Isella}, A., {Li}, H., \& {Ji}, J. 2016, \apj, 818, 76

\bibitem[{{Kanagawa} {et~al.}(2018){Kanagawa}, {Muto}, {Okuzumi}, {Tanigawa},
  {Taki}, \& {Shibaike}}]{2018ApJ...868...48K}
{Kanagawa}, K.~D., {Muto}, T., {Okuzumi}, S., {et~al.} 2018, \apj, 868, 48

\bibitem[{{Kobayashi} {et~al.}(2012){Kobayashi}, {Ormel}, \&
  {Ida}}]{2012ApJ...756...70K}
{Kobayashi}, H., {Ormel}, C.~W., \& {Ida}, S. 2012, \apj, 756, 70

\bibitem[{{Kornet} {et~al.}(2001){Kornet}, {Stepinski}, \&
  {R{\'o}{\.z}yczka}}]{2001A&A...378..180K}
{Kornet}, K., {Stepinski}, T.~F., \& {R{\'o}{\.z}yczka}, M. 2001, \aap, 378,
  180

\bibitem[{{Krijt} {et~al.}(2016){Krijt}, {Ormel}, {Dominik}, \&
  {Tielens}}]{2016A&A...586A..20K}
{Krijt}, S., {Ormel}, C.~W., {Dominik}, C., \& {Tielens}, A.~G.~G.~M. 2016,
  \aap, 586, A20

\bibitem[{{Krijt} {et~al.}(2018){Krijt}, {Schwarz}, {Bergin}, \&
  {Ciesla}}]{2018ApJ...864...78K}
{Krijt}, S., {Schwarz}, K.~R., {Bergin}, E.~A., \& {Ciesla}, F.~J. 2018, \apj,
  864, 78

\bibitem[{{Kruijer} {et~al.}(2017){Kruijer}, {Burkhardt}, {Budde}, \&
  {Kleine}}]{2017PNAS..114.6712K}
{Kruijer}, T.~S., {Burkhardt}, C., {Budde}, G., \& {Kleine}, T. 2017,
  Proceedings of the National Academy of Science, 114, 6712

\bibitem[{{Lambrechts} {et~al.}(2014){Lambrechts}, {Johansen}, \&
  {Morbidelli}}]{2014A&A...572A..35L}
{Lambrechts}, M., {Johansen}, A., \& {Morbidelli}, A. 2014, \aap, 572, A35

\bibitem[{{Li} {et~al.}(2005){Li}, {Li}, {Koller}, {Wendroff}, {Liska},
  {Orban}, {Liang}, \& {Lin}}]{2005ApJ...624.1003L}
{Li}, H., {Li}, S., {Koller}, J., {et~al.} 2005, \apj, 624, 1003

\bibitem[{{Li} {et~al.}(2009){Li}, {Lubow}, {Li}, \&
  {Lin}}]{2009ApJ...690L..52L}
{Li}, H., {Lubow}, S.~H., {Li}, S., \& {Lin}, D.~N.~C. 2009, \apjl, 690, L52

\bibitem[{{Li}(2017)}]{2017AIPC.1863X0004L}
{Li}, S. 2017, in American Institute of Physics Conference Series, Vol. 1863,
  500004

\bibitem[{{Li} \& {Li}(2014)}]{2014ASPC..488...96L}
{Li}, S., \& {Li}, H. 2014, in Astronomical Society of the Pacific Conference
  Series, Vol. 488, 8th International Conference of Numerical Modeling of Space
  Plasma Flows (ASTRONUM 2013), ed. N.~V. {Pogorelov}, E.~{Audit}, \& G.~P.
  {Zank}, 96

\bibitem[{{Li} {et~al.}(2019){Li}, {Li}, {Ricci}, {Li}, {Birnstiel}, {Isella},
  {Ansdell}, {Yuan}, {Dra̧{\.z}kowska}, \& {Stammler}}]{2019ApJ...878...39L}
{Li}, Y.-P., {Li}, H., {Ricci}, L., {et~al.} 2019, \apj, 878, 39

\bibitem[{{Lorek} {et~al.}(2018){Lorek}, {Lacerda}, \&
  {Blum}}]{2018A&A...611A..18L}
{Lorek}, S., {Lacerda}, P., \& {Blum}, J. 2018, \aap, 611, A18

\bibitem[{{Lubow} \& {D'Angelo}(2006)}]{2006ApJ...641..526L}
{Lubow}, S.~H., \& {D'Angelo}, G. 2006, \apj, 641, 526

\bibitem[{{Lyra} {et~al.}(2018){Lyra}, {Raettig}, \&
  {Klahr}}]{2018RNAAS...2d.195L}
{Lyra}, W., {Raettig}, N., \& {Klahr}, H. 2018, Research Notes of the American
  Astronomical Society, 2, 195

\bibitem[{{Mathis} {et~al.}(1977){Mathis}, {Rumpl}, \&
  {Nordsieck}}]{1977ApJ...217..425M}
{Mathis}, J.~S., {Rumpl}, W., \& {Nordsieck}, K.~H. 1977, \apj, 217, 425

\bibitem[{{Meheut} {et~al.}(2012){Meheut}, {Meliani}, {Varniere}, \&
  {Benz}}]{2012A&A...545A.134M}
{Meheut}, H., {Meliani}, Z., {Varniere}, P., \& {Benz}, W. 2012, \aap, 545,
  A134

\bibitem[{{Miranda} {et~al.}(2017){Miranda}, {Li}, {Li}, \&
  {Jin}}]{2017ApJ...835..118M}
{Miranda}, R., {Li}, H., {Li}, S., \& {Jin}, S. 2017, \apj, 835, 118

\bibitem[{{Morbidelli} {et~al.}(2016){Morbidelli}, {Bitsch}, {Crida},
  {Gounelle}, {Guillot}, {Jacobson}, {Johansen}, {Lambrechts}, \&
  {Lega}}]{2016Icar..267..368M}
{Morbidelli}, A., {Bitsch}, B., {Crida}, A., {et~al.} 2016, \icarus, 267, 368

\bibitem[{{Nakagawa} {et~al.}(1986){Nakagawa}, {Sekiya}, \&
  {Hayashi}}]{1986Icar...67..375N}
{Nakagawa}, Y., {Sekiya}, M., \& {Hayashi}, C. 1986, Icarus, 67, 375

\bibitem[{{Okuzumi} {et~al.}(2012){Okuzumi}, {Tanaka}, {Kobayashi}, \&
  {Wada}}]{2012ApJ...752..106O}
{Okuzumi}, S., {Tanaka}, H., {Kobayashi}, H., \& {Wada}, K. 2012, \apj, 752,
  106

\bibitem[{{Ormel} {et~al.}(2007){Ormel}, {Spaans}, \&
  {Tielens}}]{2007A&A...461..215O}
{Ormel}, C.~W., {Spaans}, M., \& {Tielens}, A.~G.~G.~M. 2007, \aap, 461, 215

\bibitem[{{Paardekooper} \& {Mellema}(2004)}]{2004A&A...425L...9P}
{Paardekooper}, S.-J., \& {Mellema}, G. 2004, \aap, 425, L9

\bibitem[{{Paardekooper} \& {Mellema}(2006)}]{2006A&A...453.1129P}
{Paardekooper}, S.~J., \& {Mellema}, G. 2006, \aap, 453, 1129

\bibitem[{{Picogna} \& {Kley}(2015)}]{2015A&A...584A.110P}
{Picogna}, G., \& {Kley}, W. 2015, \aap, 584, A110

\bibitem[{{Pinilla} {et~al.}(2012){Pinilla}, {Benisty}, \&
  {Birnstiel}}]{2012A&A...545A..81P}
{Pinilla}, P., {Benisty}, M., \& {Birnstiel}, T. 2012, \aap, 545, A81

\bibitem[{{Pohl} {et~al.}(2015){Pohl}, {Pinilla}, {Benisty}, {Ataiee},
  {Juh{\'a}sz}, {Dullemond}, {Van Boekel}, \& {Henning}}]{2015MNRAS.453.1768P}
{Pohl}, A., {Pinilla}, P., {Benisty}, M., {et~al.} 2015, \mnras, 453, 1768

\bibitem[{{Raettig} {et~al.}(2015){Raettig}, {Klahr}, \&
  {Lyra}}]{2015ApJ...804...35R}
{Raettig}, N., {Klahr}, H., \& {Lyra}, W. 2015, \apj, 804, 35

\bibitem[{{Rice} {et~al.}(2006){Rice}, {Armitage}, {Wood}, \&
  {Lodato}}]{2006MNRAS.373.1619R}
{Rice}, W.~K.~M., {Armitage}, P.~J., {Wood}, K., \& {Lodato}, G. 2006, \mnras,
  373, 1619

\bibitem[{{Ruge} {et~al.}(2016){Ruge}, {Flock}, {Wolf}, {Dzyurkevich},
  {Fromang}, {Henning}, {Klahr}, \& {Meheut}}]{2016A&A...590A..17R}
{Ruge}, J.~P., {Flock}, M., {Wolf}, S., {et~al.} 2016, \aap, 590, A17

\bibitem[{{Schaffer} {et~al.}(2018){Schaffer}, {Yang}, \&
  {Johansen}}]{2018A&A...618A..75S}
{Schaffer}, N., {Yang}, C.-C., \& {Johansen}, A. 2018, \aap, 618, A75

\bibitem[{{Sengupta} {et~al.}(2019){Sengupta}, {Dodson-Robinson}, {Hasegawa},
  \& {Turner}}]{2019ApJ...874...26S}
{Sengupta}, D., {Dodson-Robinson}, S.~E., {Hasegawa}, Y., \& {Turner}, N.~J.
  2019, \apj, 874, 26

\bibitem[{{Shakura} \& {Sunyaev}(1973)}]{1973A&A....24..337S}
{Shakura}, N.~I., \& {Sunyaev}, R.~A. 1973, \aap, 24, 337

\bibitem[{{Shibaike} {et~al.}(2017){Shibaike}, {Okuzumi}, {Sasaki}, \&
  {Ida}}]{2017ApJ...846...81S}
{Shibaike}, Y., {Okuzumi}, S., {Sasaki}, T., \& {Ida}, S. 2017, \apj, 846, 81

\bibitem[{{Surville} \& {Barge}(2015)}]{2015A&A...579A.100S}
{Surville}, C., \& {Barge}, P. 2015, \aap, 579, A100

\bibitem[{{Surville} {et~al.}(2016){Surville}, {Mayer}, \&
  {Lin}}]{2016ApJ...831...82S}
{Surville}, C., {Mayer}, L., \& {Lin}, D.~N.~C. 2016, \apj, 831, 82

\bibitem[{{Szul{\'a}gyi}(2017)}]{2017ApJ...842..103S}
{Szul{\'a}gyi}, J. 2017, \apj, 842, 103

\bibitem[{{Tamfal} {et~al.}(2018){Tamfal}, {Dr{\c a}{\.z}kowska}, {Mayer}, \&
  {Surville}}]{2018ApJ...863...97T}
{Tamfal}, T., {Dr{\c a}{\.z}kowska}, J., {Mayer}, L., \& {Surville}, C. 2018,
  \apj, 863, 97

\bibitem[{{Vorobyov} {et~al.}(2018){Vorobyov}, {Akimkin}, {Stoyanovskaya},
  {Pavlyuchenkov}, \& {Liu}}]{2018A&A...614A..98V}
{Vorobyov}, E.~I., {Akimkin}, V., {Stoyanovskaya}, O., {Pavlyuchenkov}, Y., \&
  {Liu}, H.~B. 2018, \aap, 614, A98

\bibitem[{{Weber} {et~al.}(2018){Weber}, {Ben{\'{\i}}tez-Llambay}, {Gressel},
  {Krapp}, \& {Pessah}}]{2018ApJ...854..153W}
{Weber}, P., {Ben{\'{\i}}tez-Llambay}, P., {Gressel}, O., {Krapp}, L., \&
  {Pessah}, M.~E. 2018, \apj, 854, 153

\bibitem[{{Weidenschilling}(1977)}]{1977Ap&SS..51..153W}
{Weidenschilling}, S.~J. 1977, \apss, 51, 153

\bibitem[{{Windmark} {et~al.}(2012){Windmark}, {Birnstiel}, {G{\"u}ttler},
  {Blum}, {Dullemond}, \& {Henning}}]{2012A&A...540A..73W}
{Windmark}, F., {Birnstiel}, T., {G{\"u}ttler}, C., {et~al.} 2012, \aap, 540,
  A73

\bibitem[{{Xu} {et~al.}(2017){Xu}, {Bai}, \&
  {Murray-Clay}}]{2017ApJ...847...52X}
{Xu}, Z., {Bai}, X.-N., \& {Murray-Clay}, R.~A. 2017, \apj, 847, 52

\bibitem[{{Youdin} \& {Lithwick}(2007)}]{2007Icar..192..588Y}
{Youdin}, A.~N., \& {Lithwick}, Y. 2007, Icarus, 192, 588

\bibitem[{{Zhu} {et~al.}(2012){Zhu}, {Nelson}, {Dong}, {Espaillat}, \&
  {Hartmann}}]{2012ApJ...755....6Z}
{Zhu}, Z., {Nelson}, R.~P., {Dong}, R., {Espaillat}, C., \& {Hartmann}, L.
  2012, \apj, 755, 6

\bibitem[{{Zhu} {et~al.}(2014){Zhu}, {Stone}, {Rafikov}, \&
  {Bai}}]{2014ApJ...785..122Z}
{Zhu}, Z., {Stone}, J.~M., {Rafikov}, R.~R., \& {Bai}, X.-n. 2014, \apj, 785,
  122

\bibitem[{{Zsom} \& {Dullemond}(2008)}]{2008A&A...489..931Z}
{Zsom}, A., \& {Dullemond}, C.~P. 2008, \aap, 489, 931

\bibitem[{{Zsom} {et~al.}(2011){Zsom}, {Ormel}, {Dullemond}, \&
  {Henning}}]{2011A&A...534A..73Z}
{Zsom}, A., {Ormel}, C.~W., {Dullemond}, C.~P., \& {Henning}, T. 2011, \aap,
  534, A73

\end{thebibliography}



\end{document}